\documentclass[twocolumn]{aastex6}

\usepackage{amsfonts}
\usepackage{amsmath}
\usepackage{amssymb}
\usepackage{graphicx}
\usepackage{color}
\usepackage{natbib}

\def\aap{A\&A}

\def\apjl{ApJ}

\def\apjs{ApJS}

\def\aap{A\&A}

\def\apjl{ApJ}

\def\apjs{ApJS}

\newif\ifAMStwofonts
\AMStwofontstrue

\graphicspath{{./fig/}}

\shorttitle{Magnetic activity of G-type main sequence \emph{Kepler} targets}
\shortauthors{Mehrabi et al.}

\begin{document}

\title{Magnetic activity analysis for a sample of G-type main sequence \emph{Kepler} targets}

\author{Ahmad Mehrabi\altaffilmark{1}}
\affil{ Department of Physics, Bu Ali Sina University,65178, 016016, Hamedan, Iran\\
School of Astronomy, Institute for Research in Fundamental Sciences (IPM), 19395-5531,  Tehran, Iran }

\author{Han He\altaffilmark{2}}
\affil{National Astronomical Observatories, Chinese Academy of Sciences, Beijing, China\\
Key Laboratory of Solar Activity, Chinese Academy of Sciences, Beijing, China}

\and

\author{Habib khosroshahi\altaffilmark{3}}
\affil{School of Astronomy, Institute for Research in Fundamental Sciences (IPM), 19395-5531,  Tehran, Iran }

\altaffiltext{1}{mehrabi@basu.ac.ir}





\begin{abstract}
The variation of a stellar light curve owing to the rotational modulation by the magnetic features (starspots and faculae) on the star's surface can be used to investigate the magnetic properties of the host star. In this paper, we use the periodicity and magnitude of the light-curve variation, as two proxies, suggested by \citep{2015ApJS..221...18H}, to study the stellar magnetic properties for a large sample of G-type main sequence \emph{Kepler} targets, for which the rotation periods recently determined by \citep{2014ApJS..211...24M}. By analyzing the correlation between the two magnetic proxies, it is found that: (1) The two proxies are positively correlated for most of the stars in our sample, and the percentages of negative, zero, and positive correlation are $4.27\%$, $6.81\%$, and $88.91\%$, respectively; (2) Negative correlation stars cannot have large magnitude of light-curve variation; (3) with the increase of rotation period, the relative number of positive correlation stars decreases and the negative correlation one increases.  These results indicate that the stars with shorter rotation period tend to have positive correlation between the two proxies, and a good portion of the positive correlation stars have larger magnitude of light-curve variation (and hence more intense magnetic activities) than the negative correlation stars.
\end{abstract}

\keywords{stars: activity -- stars: magnetic field -- stars: rotation -- stars: solar-type}

\section{Introduction}
The \emph{Kepler} mission \citep{2010Sci...327..977B} observes roughly 115 square degrees of sky in 423--897 nm wavelength band \citep{2010ApJ...713L..79K} to find possible transits of the Earth-like planets among more than $1.5\times 10^{5}$ stars. The large number of stars and the  high precision photometry of the data make the \emph{Kepler} targets a good sample for studying the properties of stars. One interesting aspect to study using \emph{Kepler} data is the magnetic activity of solar-type stars, which is believed to stem from the convective zone of solar-type stars through dynamo process \citep{2005LRSP....2....8B}. For the stars with vibrant magnetic activity, two prominent phenomena are expected to occur which could be seen in their light curves: First, inhomogeneous dark starspots and bright faculae could generate a gradual fluctuation in the light-curves by the process of rotational modulation \citep{2011A&A...529A..89D}, which might be utilized to estimate the rotation periods of stars \citep{2013A&A...560A...4R,2013A&A...557L..10N,2013MNRAS.432.1203M,2014ApJS..211...24M}; the second prominent phenomenon is stellar flare, which appears in light curves as the sudden spikes that possess a flare-type fine profile \citep{2011AJ....141...50W, 2012Natur.485..478M, 2013ApJS..209....5S, 2015MNRAS.447.2714B, 2016AcASn..57....9Y, 2016ApJ...829...23D}. As shown by \citep{2012Natur.485..478M} and \citep{2013ApJS..209....5S}, the flare spikes can only be discerned from the light curves for a small part of \emph{Kepler} stars, whereas the fluctuation owing to the rotational modulation is common in \emph{Kepler} light curves \citep{2013ApJ...771..127N, 2014ApJS..211...24M}.

Since the dark starspots and/or bright faculae which cause the light-curve modulation are associated with the magnetic fields on the surface of solar-type stars \citep{2005LRSP....2....8B, 2012LRSP....9....1R}, the fluctuation information of the modulated light-curves could be used to investigate the magnetic properties of the host stars \citep{2015ApJS..221...18H}. Observations of the Sun have demonstrated the relationship between the solar magnetic activity and the variation of the solar irradiance flux. On the Sun, the dark sunspots are originated from the intense and concentrated magnetic field regions in the photosphere, and the bright faculae are associated with the enhanced network magnetic fields which spread over a larger area \citep{2006RPPh...69..563S}. Investigations by \citep{1998ApJ...492..390L} and \citep{2000A&A...353..380F} revealed that both the sunspots and the faculae play important roles in the solar light-curve modulation: the dark sunspot regions decrease the solar irradiance while the bright facula regions increase the flux of the Sun. In particular, the overall changes of both sunspots and faculae present a long-term variation of the solar irradiance flux along with the 11-year solar cycle, and the long length scale trend of the variation is governed by faculae \citep{2004A&ARv..12..273F}.

Time evolution of the magnetic features leads to the complexity of the shapes of solar light-curve
 \citep{2003A&A...403.1135L, 2015ApJS..221...18H}. It has been found that the lifetimes of most sunspot regions are shorter than the rotation period of the Sun, while facula regions have relatively longer lifetimes owing to their larger spatial scale \citep{1985AuJPh..38..961Z,2000Natur.408..445S}. The sunspot regions with shorter lifetimes (and hence faster time evolution) could cause the irregular variations of light curve in the process of rotational modulation \citep{2003A&A...403.1135L}, and the facula regions with longer lifetimes tend to cause the periodic light curves \citep{2015ApJS..221...18H}.  That is, the degree of periodicity of a light curve reflects the stability of the magnetic features that cause the rotational modulation \citep{2015ApJS..221...18H}.Note that the periodicity of light curve discussed here (in the context of rotational modulation) is different from the long-term (11-year) cyclic variation of the solar flux \citep{2004A&ARv..12..273F} mentioned above.

  The relation between the degree of periodicity of the light curves and the evolution properties of the magnetic features can also be applied to the solar-type stars observed with \emph{Kepler} \citep{2015ApJS..221...18H}. Meanwhile, the spatial size of the magnetic features (and hence the intensity level of stellar magnetic activities) can be reflected by the fluctuation magnitude of the modulated light-curves \citep{2011AJ....141...20B, 2013ApJ...769...37B}. \citep{2015ApJS..221...18H} suggested the periodicity and the magnitude of light-curve variation as two independent magnetic proxies for solar-type stars. They also propose two measures, $i_{\rm AC}$ and $R_{\rm eff}$ , to quantitatively describe these two proxies.

In the paper by \citep{2015ApJS..221...18H}, they applied the two magnetic proxies to the Sun (rotation period: 25.38 days) and two solar-type stars observed with \emph{Kepler}. One \emph{Kepler} star (KIC 9766237) has a moderate rotation period (14.7days) and another \emph{Kepler} star (KIC 10864581) has a smaller period (6 days). Their analysis shows that: for the solar-type star with smaller rotation period, the time variations of the two proxies present positive correlation; for the Sun with longer rotation period, the two proxies are negatively correlated; and for the star with moderate period, the two proxies show weak correlation.  These results demonstrate a distinct variety of the magnetic activity properties on the different solar-type stars, and provide a clue to the relationship between the correlations of the two magnetic proxies and the rotation periods of solar-type stars.

In this paper, we extend the work of \citep{2015ApJS..221...18H} and use the two magnetic proxies to study the statistical properties of the magnetic activities for a large sample of G-type main sequence stars in \emph{Kepler} targets. Section (\ref{sec2}) gives a brief description of the two light-curve-based magnetic proxies. Section (\ref{sec3}) explains the criteria for the \emph{Kepler} targets selection and all steps we proceed for processing the data. In section (\ref{sec4}) we present the results of the analysis of magnetic activities as well as the properties of the stars which are used in our analysis. Finally in section (\ref{sec_concl}) we conclude and discuss about the important aspects of our results.

\section{periodicity and magnitude of light-curve variation}\label{sec2}
The periodicity of a time series can be quantitatively measured by computing the autocorrelation function of the time series. That is, a periodic time series repeats itself after a certain time lag and this self similarity increases the correlation coefficient between the original time series and the time series with the time lag corresponding to period of it. So the autocorrelation algorithm could be used to measure the periodicity of a time series \citep{2015ApJS..221...18H}.

For an observed time series or light-curve as $\{X_{t},t=0,1,2,3,...,N-1\}$ with $N$ points, the autocorrelation function (ACF) is defined via:
\begin{equation}\label{eq:rho}
\rho(h)=\frac{\sum_{t=0}^{t=N-1-h}(X_{t+h}-\bar{X})(X_{t}-\bar{X})}{\sum_{t=0}^{t=N-1}(X_{t}-\bar{X})^2},
\end{equation}
where $h$ is the time lag and $\bar{X}$ is the mean value of the time series.
For a perfect periodic time series, the fluctuation range of $\rho(h)$ are large and $\rho(h)$ has maximum (minimum) values at each full period (half period) time lag. In contrast, for a random time series without any periodic feature, the derived autocorrelation function has tiny fluctuations without any regular patterns. To find more details regarding the autocorrelation functions for a perfect periodic time series and a random one, see \citep{2015ApJS..221...18H}. Moreover, as mentioned above, the autocorrelation function for a periodic time series fluctuates with the same period as the original time series. This opens a window to find periods of stars by computing the autocorrelation functions of their light curves. In fact, \citep{2013MNRAS.432.1203M,2014ApJS..211...24M} used the autocorrelation method to measure the rotation periods for a large sample of \emph{Kepler} stars.

As proposed by \citep{2015ApJS..221...18H}, one can measure the degree of periodicity of a light curve by computing the average value of $|\rho(h)|$ over the interval $0<h<N/2$:
\begin{equation}\label{eq:iac}
i_{\rm AC}=\frac{2}{N}\int_{0}^{N/2}|\rho(h)| dh.
\end{equation}
Since this quantity reflects the total area of the region enclosed by the autocorrelation function, it is large for a periodic time series rather than a random time series. The interval of the integration in equation (\ref{eq:iac}) is limited to $N/2$ because if $h > N/2$, not all data points in the light curve are utilized to calculate $\rho(h)$ as defined in equation (\ref{eq:rho}) and this portion of $\rho(h>N/2)$ can not reflect the whole property of the light curve. For a perfect sine curve $x=A\sin(\omega t)$ with total length $L$, the autocorrelation function is roughly given by $\rho(h)=(1-\frac{h}{L})\cos(\omega h)$ which results $i_{\rm AC}\approx 0.48$. So generally the quantity $i_{\rm AC}$ varies between $(0,0.48)$ and might be considered as a quantitative measure to show the degree of periodicity of a given time series. For a time series with irregular variation, the $i_{\rm AC}$ value is relatively small, in contrast with a time series with stable periodic fluctuation which has a relatively higher value of $i_{\rm AC}$.

For a given stellar light-curve, fluctuation amplitude is not generally uniform. So one might define an effective range of fluctuation $R_{\rm eff}$ to describe the magnitude of light-curve variation quantitatively. As adopted by \citep{2011AJ....141...20B, 2013ApJ...769...37B}, fluctuation range of a light-curve is the distance between the crest and trough of the light curve. It is convenient to normalize the light-curve data before defining the measure of fluctuation magnitude. One can normalize a light curve $\{X_{t},t=0,1,2,3,...,N-1\}$ to its median value as:
\begin{equation}
x_{t}=\frac{X_{t}-\tilde{X}}{\tilde{X}},
\end{equation}
 where $\tilde{X}$ is the median of the light curve. In this case, $x_{t}$ fluctuates around the zero and is more convenient for further analyses. Note that both $X_{t}$ and $x_{t}$ yield the same values of $\rho(h)$ and $i_{\rm AC}$. Finally, the effective fluctuation range is given by
 \begin{equation}\label{eq:reff}
 R_{\rm eff}=2(\sqrt{2}x_{\rm rms}),
 \end{equation}
 where $x_{\rm rms}$ is the rms value of $x_{t}$ with definition \citep{2010Sci...329.1032G, 2011ApJ...732L...5C},
 \begin{equation}
 x_{\rm rms}=\sqrt{\frac{1}{N}\sum_{t=0}^{t=N-1}x_{t}^2}.
 \end{equation}
For a perfect sine wave $x=A\sin{\omega t}$, $x_{\rm rms}$ is $\frac{A}{\sqrt{2}}$ so we have $R_{\rm eff}=2(\sqrt{2}\frac{A}{\sqrt{2}})=2A$ which is the true fluctuation range (distance between crest and trough) of the light curve. In fact the factor $2\sqrt{2}$ in equation (\ref{eq:reff}) is put to give a corrected value of the fluctuation range of a light curve (for an illustrative example, see \citep{2015ApJS..221...18H}).

For a real stellar light curve the effective range of fluctuation $R_{\rm eff}$ is a proxy of magnetic activity which reveals the size of the magnetic features on the surface of the star. On the other hand, another proxy $i_{\rm AC}$ could be used to measure the stability of magnetic features. The correlation of these quantities might give us important information about magnetic activity properties of stars and could be useful for magnetic activity analyses and theory of stellar dynamo. In this work we select a sample of G-type main sequence \emph{Kepler} targets to investigate their magnetic activities. After describing our targets selection and their stellar properties we will present our analysis results.

\section{\emph{Kepler} targets selection and data processing}\label{sec3}
In our analysis, we use the light curves of \emph{Kepler} targets obtained by the photometry instrument in long-cadence (LC) mode, i.e., one data point in every 29.4 minutes \citep{2010ApJ...713L.120J}. The systematic errors are corrected by passing the raw data to the Presearch Data Conditioning (PDC) module \citep{2012PASP..124.1000S, 2012PASP..124..985S} of the \emph{Kepler} data processing pipeline. This procedure yields the PDC flux data. Note that the PDC module removes systematic errors while keeping astrophysical signals, which makes the PDC flux data be appropriate for the astrophysical analyses. In this work, we use the PDC flux data in \emph{Kepler} Data Release 25 \citep{Release25} for our analysis. The data files were downloaded from the MAST (Mikulski Archive for Space Telescopes) web site \footnote{https://archive.stsci.edu/kepler/publiclightcurves.html}.

The light-curve data of \emph{Kepler} are divided into 18 quarters (Q0--Q17). Each quarter (except Q0, Q1, and Q17) contains about three months of continuous data \citep{2010ApJ...713L.115H}. Because the \emph{Kepler} telescope rolls $90^{\circ}$ about its axis between successive quarters, the data of a same star are obtained with different CCD modules \citep{2010ApJ...713L..79K}, which leads to the discontinuities of \emph{Kepler} light curves at the end of each quarter. Due to this, we process the light-curve data of each quarter separately.

The step by step procedures for \emph{Kepler} targets selection and data processing in this work are described as follows.

\begin{enumerate}
	
	\item Since we aim at studying the magnetic properties of solar-type stars in \emph{Kepler} field, we should select those stars with confirmed magnetic activity. To do this,  we use the \emph{Kepler} targets in the catalog of \citep{2014ApJS..211...24M} which contains about 34000 main sequence stars with confirmed rotation period.  As mentioned above, the rotation periods of these stars can be obtained owing to the existing of the inhomogeneous spots and faculae which result a fluctuating signature in the light curves through rotational modulation and indicate the existence of magnetic activity on the stars. The periods of the stars in the catalog of \citep{2014ApJS..211...24M} are obtained using the ACF method and this catalog provides the largest sample set of the main sequence stars with known rotation period.
	
	\item To select the G-type main sequence stars from the catalog of \citep{2014ApJS..211...24M},  at first, the $B-V$ color for the stars in this catalog are obtained via the relation $B-V =0.98(g-r) + 0.22$ \citep{2005AJ....130..873J}, and then, by using appendix B of \citep{Gray-2005}, the G-type stars are selected. In this step those stars with $0.59<B-V<0.74$ are identified as G-type stars, and after this filtering we are left with 6717 stars which are depicted in H-R diagram in Fig.(\ref{fig:h-r}). Note that here we use the relation $L/L_{\odot}=R^2(\frac{T_{\rm eff}}{T_{\odot}})^4$ \citep{2011AJ....142..112B} to find the luminosity values of the stars. 
	\begin{figure}
		\centering
		\includegraphics[width=0.5\textwidth]{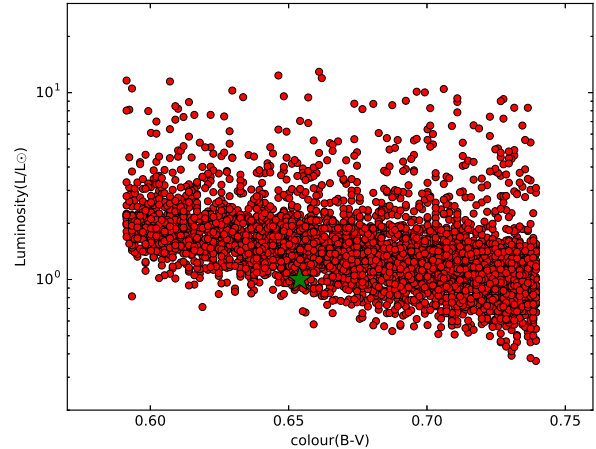}
		\caption{H-R diagram of the G-type stars selected from the catalog of \citep{2014ApJS..211...24M}. The green star symbol shows the location of the Sun for reference.}
		\label{fig:h-r}
	\end{figure}
	
	\item For fulfilling the aim of our analysis, the stars should be continuously monitored. In this step, we select those stars which have data in all quarters from Q2 to Q16, and after this filtering we are left with 5227 targets. Note that Q0, Q1, and Q17 are not full length (three months) quarters and the data in these three quarters are not adopted in this work.
	
	\item One possible source of getting wrong results is the gaps within each quarter. For this reason we neglect those stars with the gaps longer than 3 times of rotation period (the value of 3 being an empirical criterion). This filtering omits most stars with short period. Note that for those selected stars with small gaps, we use a simple linear interpolation algorithm to obtain continuous time series. Another possible option is that,one might fill any missing flux values with zero. This method was adopted in \citep{2014ApJS..211...24M} to compute ACF from a light curve, and in our case it does not change our results significantly. For example, in the extreme case, using these two methods for filling flux within gaps results a difference of  $2-3\%$ in $i_{\rm AC}$ quantity.
	
	\item Another important issue is that for a star with rotation period comparable to the length of a quarter, the quantity $i_{\rm AC}$ may not work correctly due to fewer repeated periods in the light-curve. To obtain a reliable $i_{\rm AC}$ value, a light curve should contain several periods. For this reason, we neglect those stars with rotation period longer than 30 days. Again this value is an empirical value for omitting long period stars.
	
	\item For those stars that passed above criteria (4094 stars in total), the PDC flux data in each quarter are normalized by its median value to obtain the relative flux
	\begin{equation}
	f=\frac{F-\tilde{F}}{\tilde{F}},~~~~~\tilde{F}={\rm median}(F).
	\end{equation}
	This quantity fluctuates around zero and is convenient for obtaining $R_{\rm eff}$.
	
	\item Since the transient variations in \emph{Kepler} light-curves, which consist of noises, outliers, flare spikes, and granulation driven flickers \citep{2014ApJ...781..124C,2014AA...570A..41K}, have potential to disturb the activity analysis via $(i_{\rm AC},R_{\rm eff})$ pairs, they are filtered out using a low pass Sinc filter. After performing such low pass filter for a light curve, the gradual component which is suitable for our analysis remains. However the upper cutoff frequency of the filter has to be determined empirically. In Fig.(\ref{fig:light_curve}) the original and filtered light-curves (Q2--Q5) for \emph{Kepler} target KIC 892713 with 4 different upper cutoff frequencies are shown. This \emph{Kepler} target passed our above criterion and its rotation period is 5.7 days. On one hand, with large value of upper cutoff frequency (upper left panel), the transit component still remains in the filtered data; on the other hand, with small value of cutoff frequency (bottom right panel), some gradual fluctuation features disappear (or disturbed) in the filtered light curve. In this work, we adopt the value of $\frac{1}{0.3P}$ for the cutoff frequency, where $P$ is the rotation period value given by the catalog of \citep{2014ApJS..211...24M}.
	\begin{figure*}
		\centering
		\includegraphics[width=0.48\textwidth]{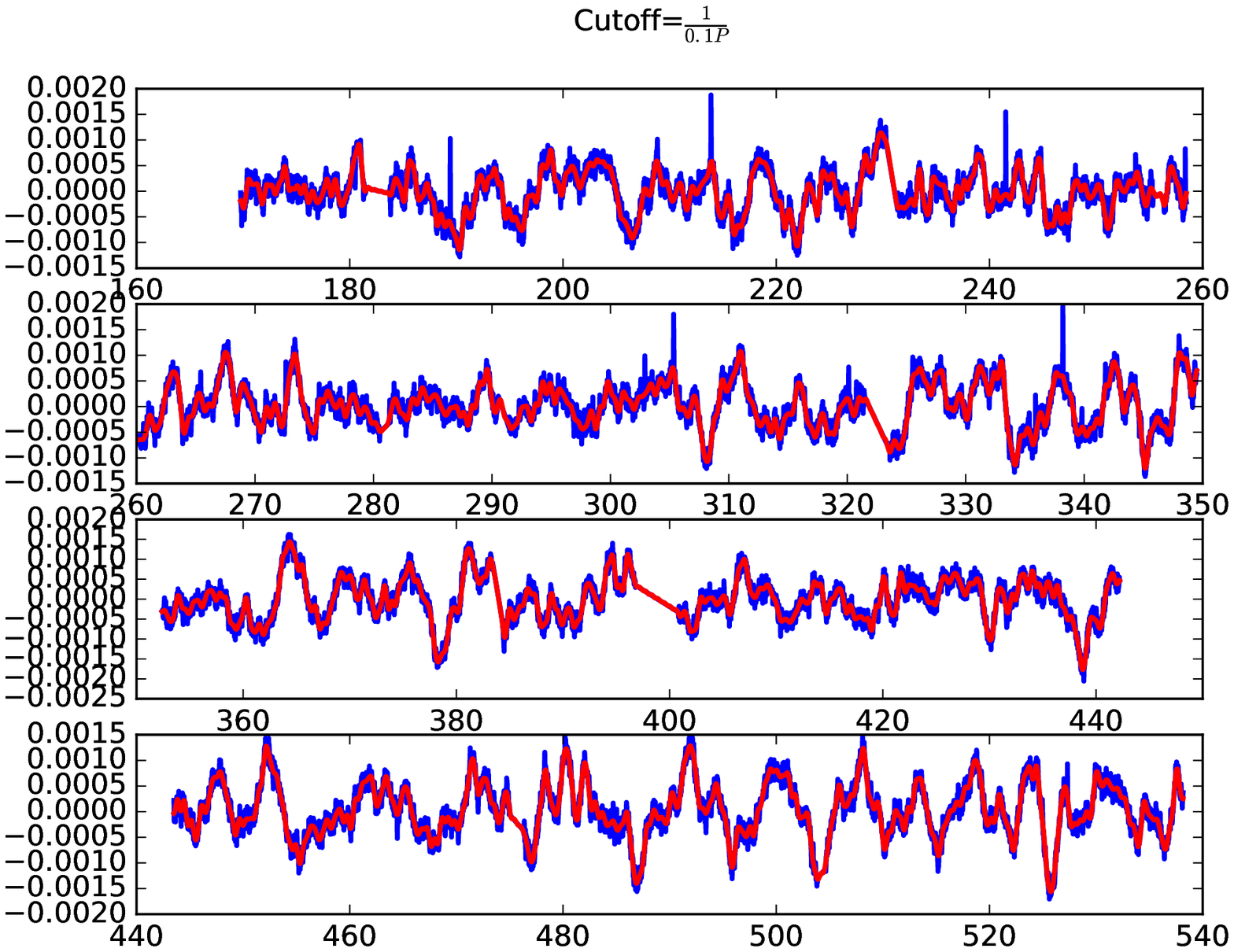}\includegraphics[width=0.48\textwidth]{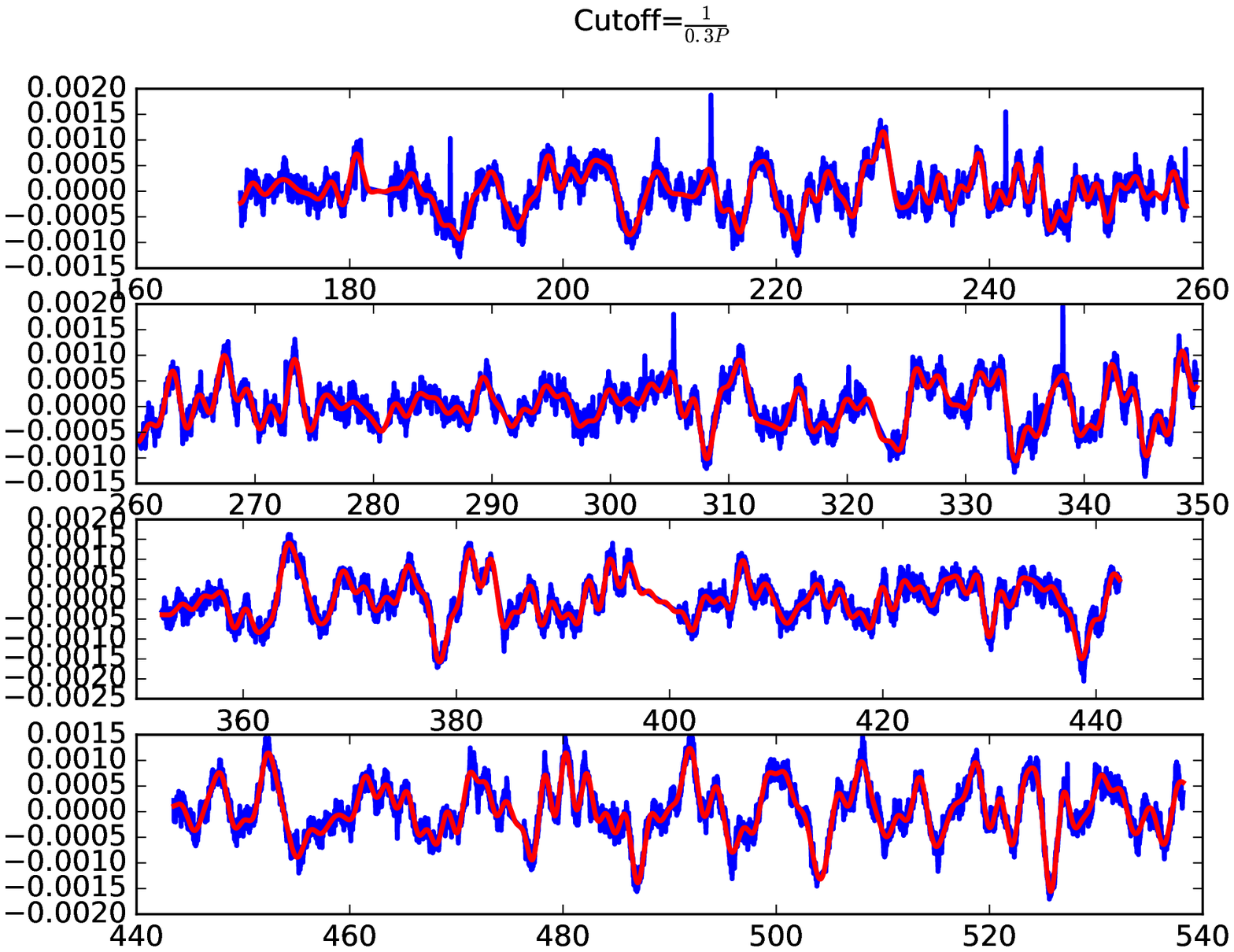}
		\includegraphics[width=0.48\textwidth]{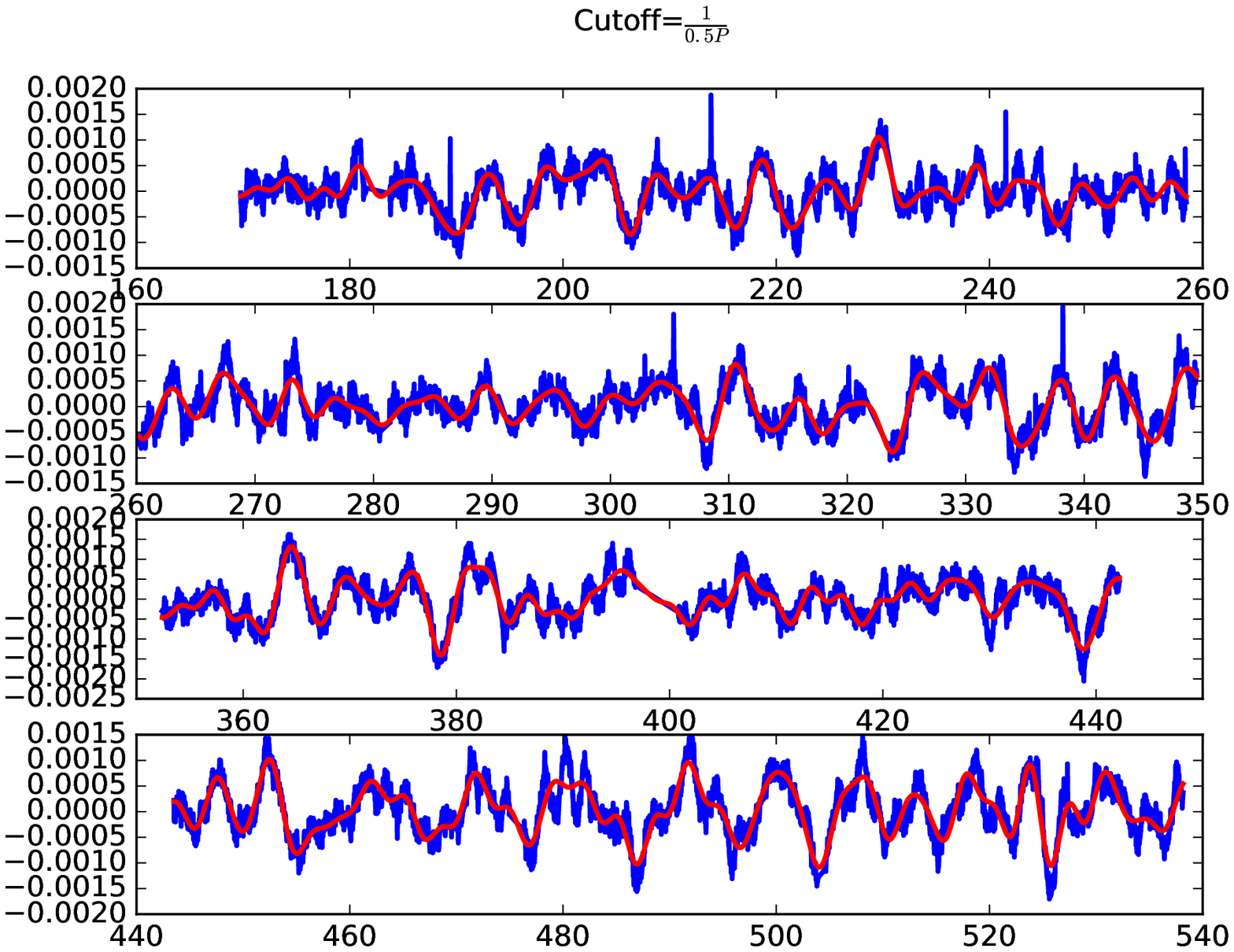}\includegraphics[width=0.48\textwidth]{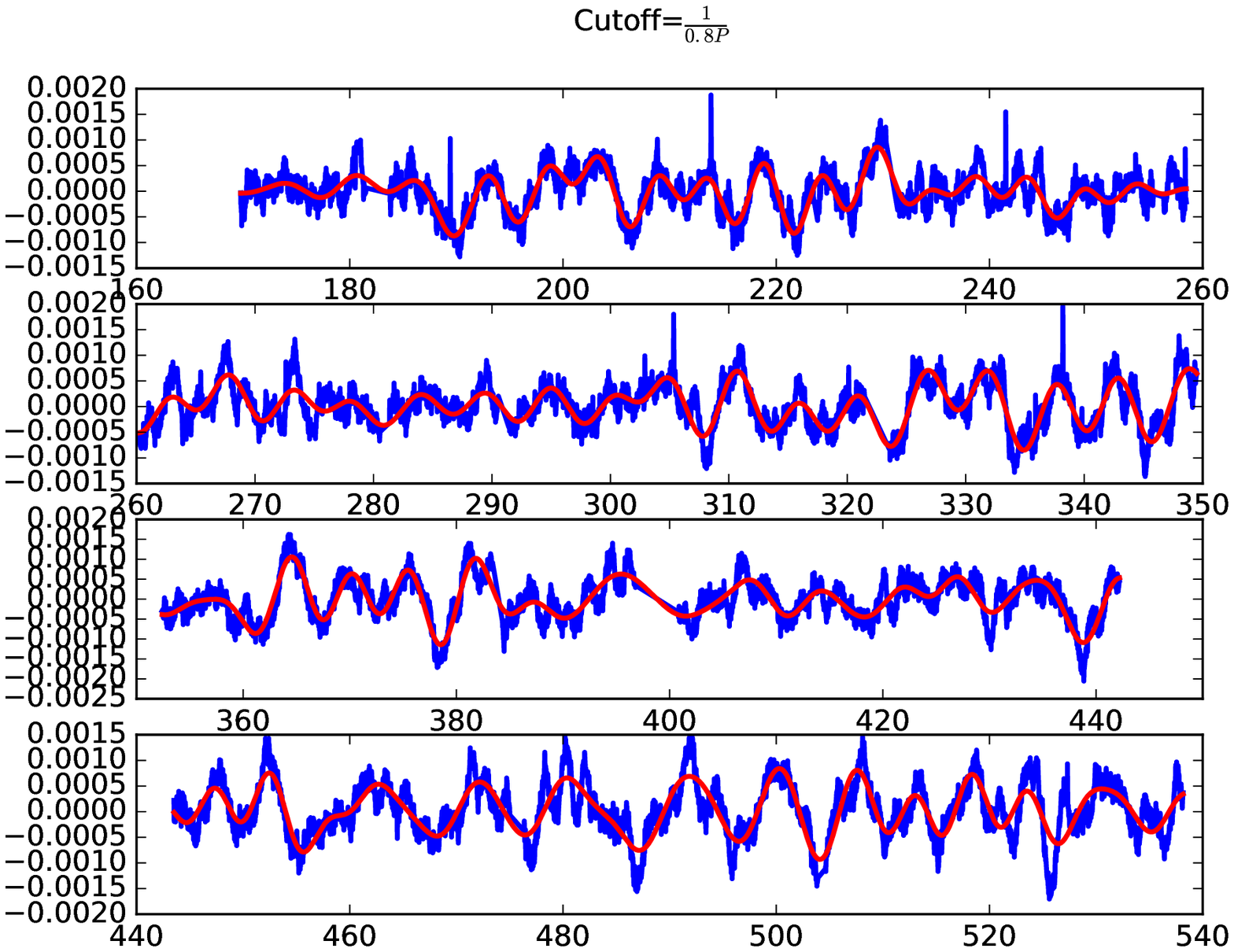}
		\caption{Original (blue color) and filtered (red color) light-curves with 4 different upper cutoff frequencies for \emph{Kepler} target KIC 892713 in Q2--Q5 (from top to bottom in each panel). The rotation period ($P$) of the star is 5.7 days. The cutoff frequencies are given at the top of each panel.}
		\label{fig:light_curve}
	\end{figure*}
	
	\item Finally, the values of the two magnetic proxies ($i_{\rm AC}$ and $R_{\rm eff}$) of each quarter from Q2 to Q16 are calculated based on the filtered light-curves for all the selected stars, and then the correlations of $(i_{\rm AC}, R_{\rm eff})$ pair (see definition in section 4.2) are calculated for each of the selected stars.
\end{enumerate}

\section{Results}\label{sec4}
\subsection{Stellar properties of the selected targets}
\begin{figure*}
	\centering
	\includegraphics[width=0.5\textwidth]{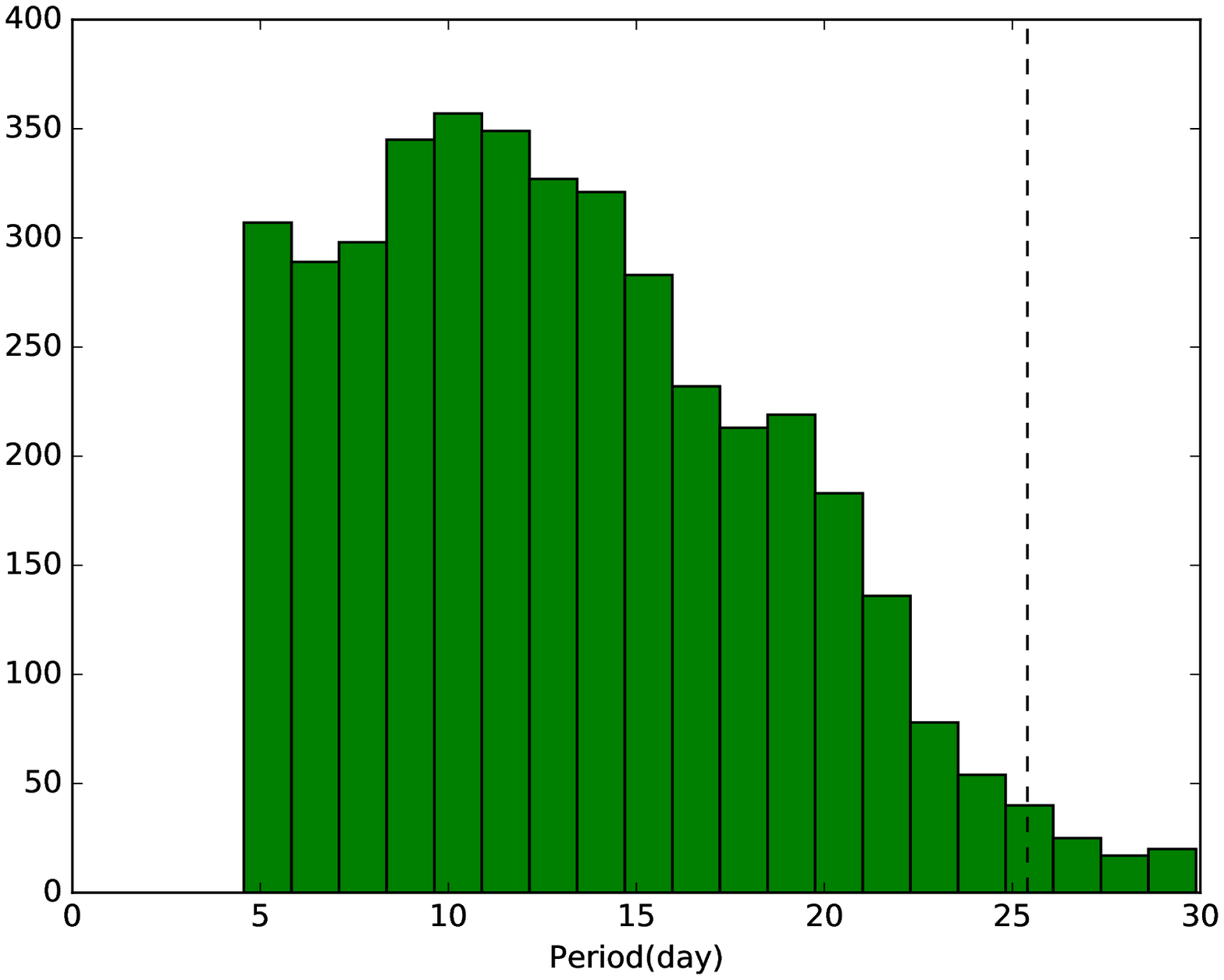}\includegraphics[width=0.5\textwidth]{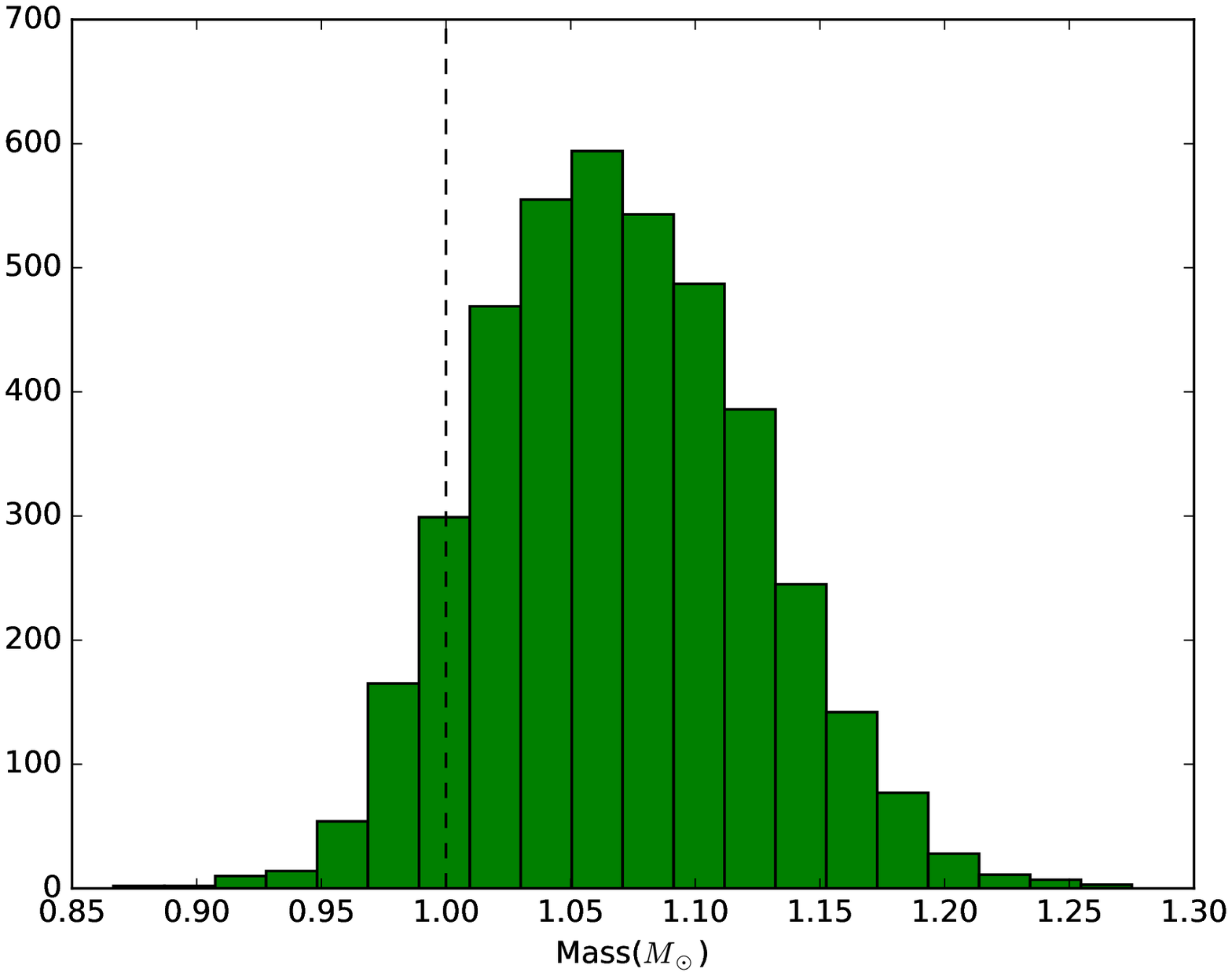}
	\includegraphics[width=0.5\textwidth]{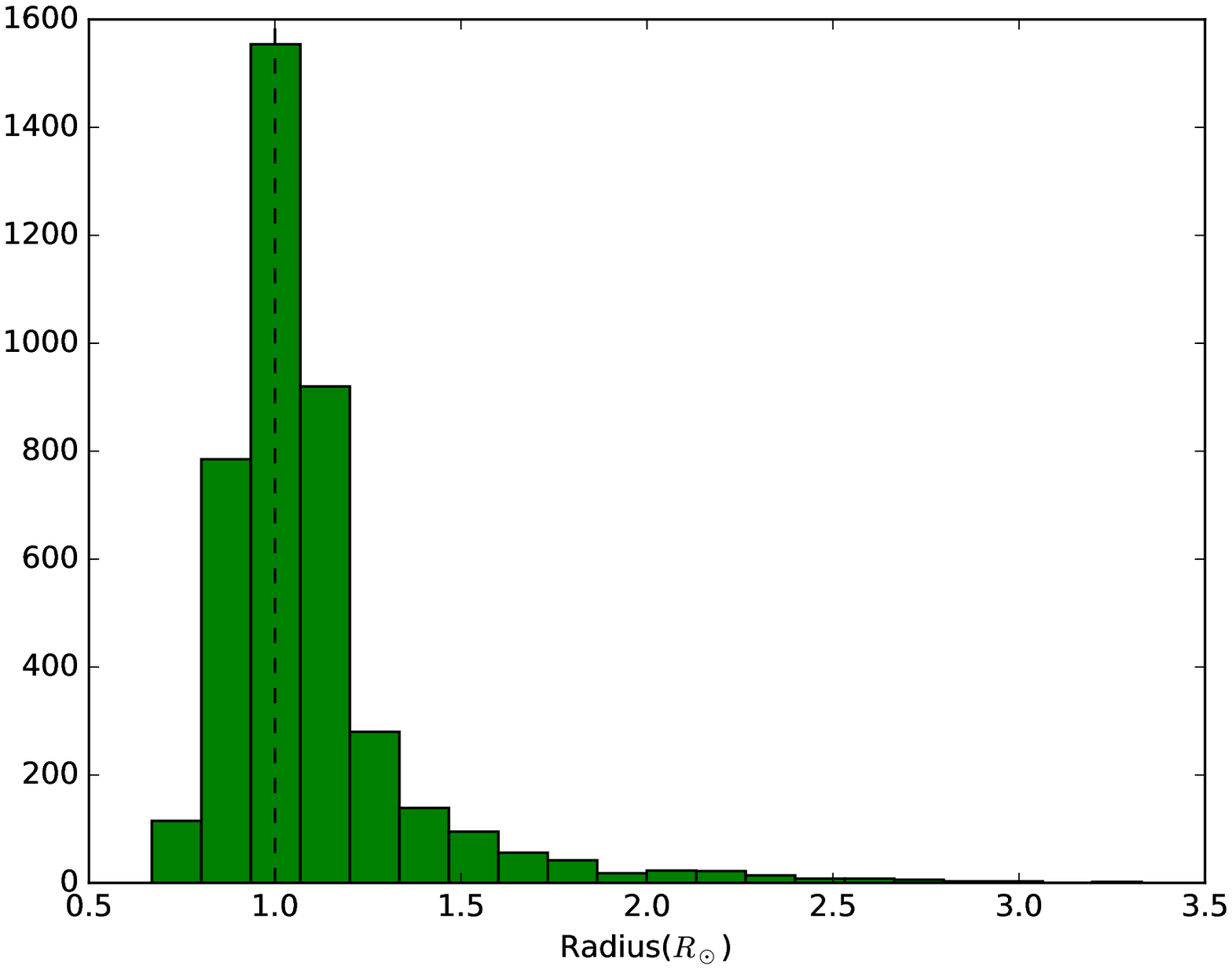}\includegraphics[width=0.5\textwidth]{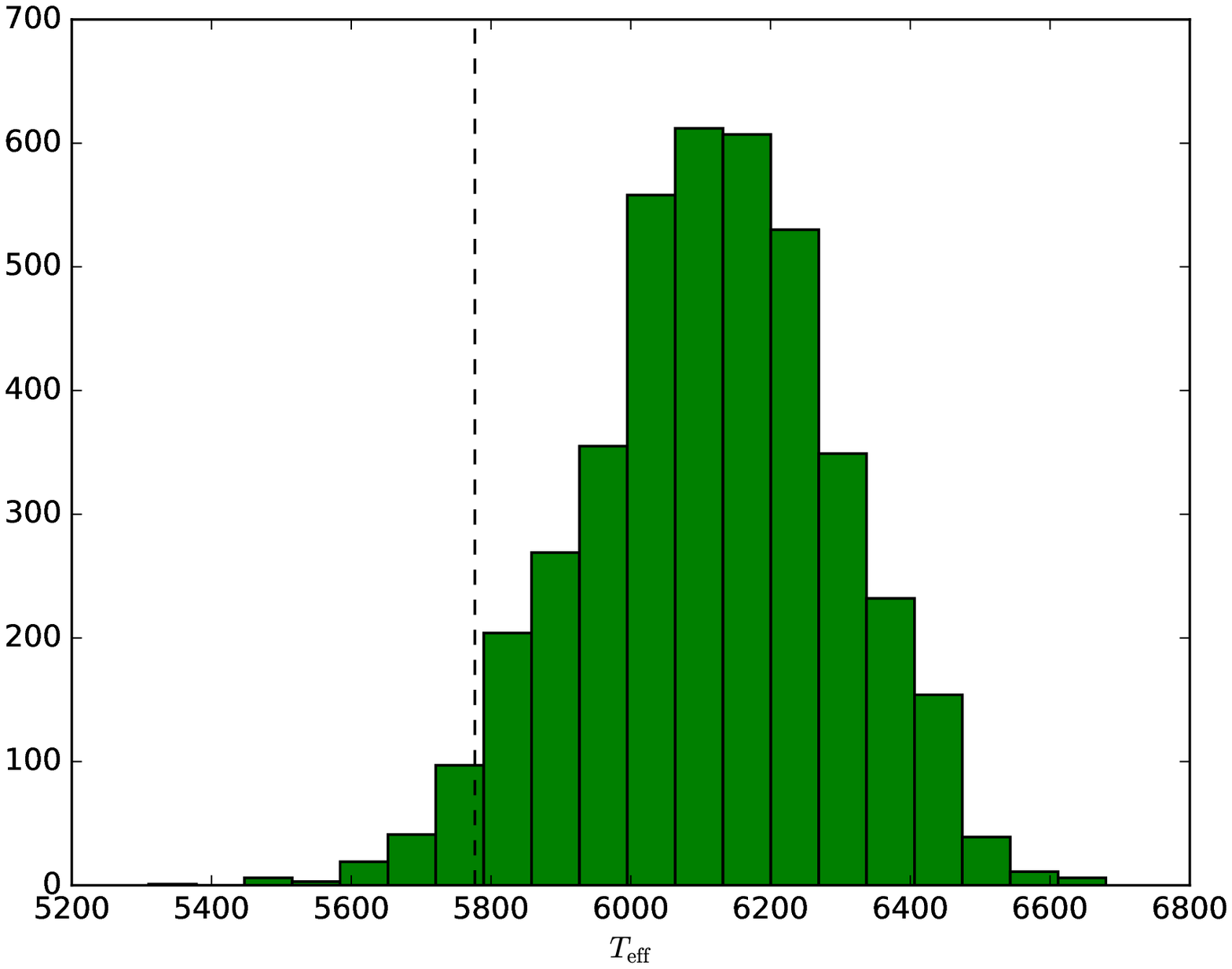}
	\caption{Distributions of rotation period, mass, radius, and effective temperature ($T_{\rm eff}$) of the selected \emph{Kepler} targets. The dashed lines indicate values of the Sun.}
	\label{fig:para_his}
\end{figure*}

Here we show the statistical analyses of the stellar properties of the selected stars. The rotation period and mass values of the stars are taken from the catalog of \citep{2014ApJS..211...24M}. Other properties like effective temperature ($T_{\rm eff}$) and radius are obtained from the \emph{Kepler} Input Catalog (KIC) \citep{2011AJ....142..112B}.
The distributions of these stellar properties are shown in Fig.(\ref{fig:para_his}).

It can be seen from Fig.(\ref{fig:para_his}) (upper left panel) that the rotation period values of the selected stars are between 5 and 30 days, which is due to our selection criteria (4) and (5) given in Section 3. The period distribution has a peak around 12 days which means most stars have medium value of period smaller than the Sun. In our catalog there are around 50 stars with period similar to that of the Sun and the ratio of slow rotating stars is relatively smaller than fast rotating stars.

In addition, the distribution of mass values shows that the stars in our sample are mostly G-type stars with the mass range 0.9--1.25 $M_{\odot}$. According to Fig.(\ref{fig:h-r}) most stars in our sample are more luminous than the Sun, so the peak value of mass distribution tend slightly toward a larger value than the Sun. In contrast to mass distribution, the radius distribution shows that most stars in our sample have the similar size of the Sun and stars with larger radius are rare in our sample. The lower right panel in Fig.(\ref{fig:para_his}) presents the effective temperature of the stars in our sample. Most stars are in the range 5600-6600 K with a peak value around $\sim 6100$K which is consistent with results in Fig.(\ref{fig:h-r}).

\subsection{Correlation of the two magnetic proxies}
The correlation (denoted by Cor) of the  $(i_{\rm AC}, R_{\rm eff})$ pair could give us useful information about the magnetic activities of the stars. $i_{\rm AC}$ is a proxy that shows the degree of periodicity of rotational modulation  which indicates the stability of the magnetic features on the surface of the star, and $R_{\rm eff}$ shows the spatial size of these features and hence intensity level of magnetic activities. The evolution of these proxies in 15 quarters (Q2-Q16), which is around 3.75 years in total, could be used to extract information about relatively long-term magnetic activity properties of these stars \citep{2015ApJS..221...18H}. The correlation between the two quantities is given by
\begin{equation}\label{eq:cross-cor}
{\rm Cor} = \frac{\sum_{q=2}^{q=16}(i_{\rm AC}^q - \bar{i}_{\rm AC})(R_{\rm eff}^q - \bar{R}_{\rm eff})}{\sqrt{\left[\sum_{q=2}^{q=16}(i_{\rm AC}^q - \bar{i}_{\rm AC})^2\right] \left[\sum_{q=2}^{q=16}(R_{\rm eff}^q - \bar{R}_{\rm eff})^2\right]}},
\end{equation}
where $\bar{i}_{\rm AC}$ and $\bar{R}_{\rm eff}$ represent the mean values of $i_{\rm AC}$ and $R_{\rm eff}$ for a star from Q2 to Q16. The correlation between the two proxies are calculated for all the selected stars in our sample. A small part of our results is presented in Table (\ref{tab:main_res}) and one can find the entire information of this table in a machine-readable form in the online journal.

Before investigating the statistical results of Cor, it is worthwhile to mention the possible sources of error in evaluating $i_{\rm AC}$ and $R_{\rm eff}$ which could result a error in Cor. There are two sources that might cause uncertainties in these parameters (apart from uncertainty that might happen in filling gaps in each quarter): (1) uncertainty in the original flux data due to systematic and stochastic errors and (2) uncertainty due to the low pass filtering process. To clarify these uncertainties, we perform a simple simulation to obtain orders of magnitude of the uncertainties. For the first case, the flux data are simulated by considering error bar in each data point with Gaussian distribution, and the standard deviations of the two proxies are considered as the uncertainties in this case. Note that in this procedure we fix the cutoff frequency to $\frac{1}{0.3P}$. Our result shows that the relative uncertainties in both proxies are of the order of $0.1\%$ or less. So the former source of error, produces very tiny uncertainty owing to very high precision of \emph{Kepler} data. For the latter case, we calculate the two proxies (for a large number of light-curves) with cutoff frequency in the range $(\frac{1}{0.05P},\frac{1}{0.65P})$ and obtain standard deviation in each proxy to find the uncertainty in this case. Our result shows that the typical relative uncertainty in $i_{\rm AC}$ is of the order of $8\%$ or less and in $R_{\rm eff}$ is of the order of $4\%$ or less. Such orders of magnitude of  uncertainties in the two proxies could not generate diverse results in evaluating the correlation defined in equation (\ref{eq:cross-cor}).

\begin{figure}
	\centering
	\includegraphics[width=0.48\textwidth]{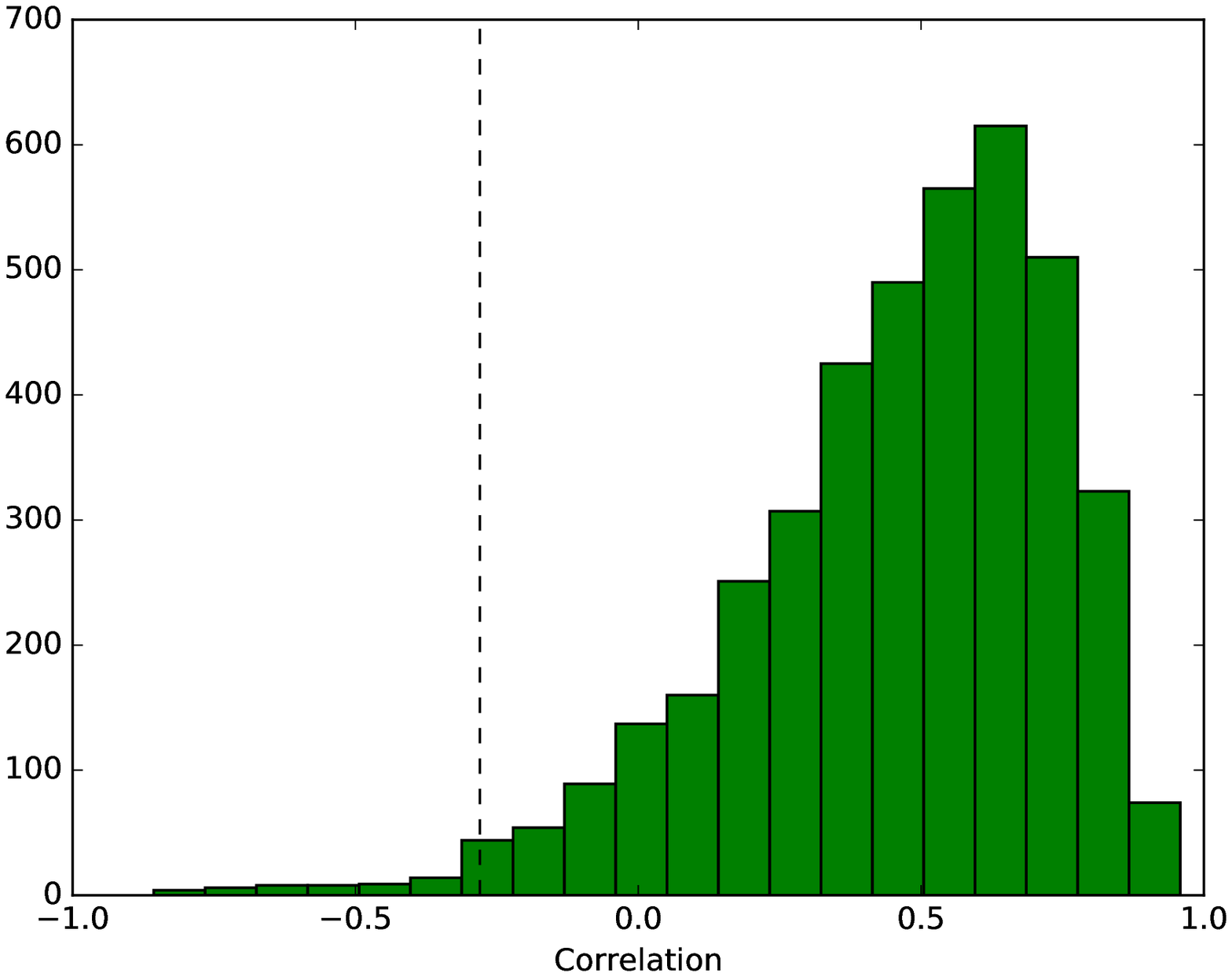}
	\caption{Distribution of correlation between $i_{\rm AC}$ and $R_{\rm eff}$.}
	\label{fig:cor_his}
\end{figure}

In Fig.(\ref{fig:cor_his}), the distribution of correlation between $i_{\rm AC}$ and $R_{\rm eff}$ for all the selected stars is shown. The correlation between the two magnetic proxies are in the range $(-1,1)$ with ${\rm Cor}=-1$ fully anticorrelated and ${\rm Cor}=1$ fully correlated. One can roughly divides the correlation values into 3 parts as $-1<{\rm Cor}<-0.1$ (negative correlation), $-0.1<{\rm Cor}<0.1$ (zero correlation), and $0.1<{\rm Cor}<1$ (positive correlation). Our analysis shows that $4.27\%$, $6.81\%$, and $88.91\%$ of the stars are in negative, zero, and positive correlation, respectively. In particular, \emph{Kepler} objects KIC 6288330 and KIC 7285775 with periods 8.88 and 8.56 days have correlations 0.958 and 0.945 which are extreme cases of positive correlation stars. On the other hand, \emph{Kepler} objects KIC 4141760 and KIC 6956092 with periods 7.54 and 7.96 days are extreme cases of negative correlation stars. In this case, we found Cor=-0.85 and Cor=-0.83 for KIC 4141760 and KIC 6956092 respectively.

\begin{table*}
	\caption{The correlation, mean values of $i_{\rm AC}$, mean value of $R_{\rm eff}$, period, mass, radius, and effective temperature of the stars presented in this work. This table is available in its entirety in a machine-readable form in the online journal.}
	\centering
	\begin{tabular}{cccccccc}
		\hline
		\hline
		KIC &  Cor & $\bar{i}_{\rm AC}$ & $\bar{R}_{\rm eff}$ & $P$ & $M$ & $R$ & $T_{\rm eff}$ \\
		    &        &                    &                     &  (days)       & ($M_{\odot})$  & ($R_{\odot}$) & (K)  \\
		\hline
		892713&0.03941&0.11185&0.00092&5.753&1.1686&2.14&6106\\
		1025986& 0.13454&0.24923&0.01805&9.724&1.0966&1.354&5841\\		
		1161620&0.49038&0.30138&0.01655&6.636 & 1.2235&1.466&6243\\	
		1162051& -0.2743&0.16340& 0.00335&19.388&1.1782&1.65&6014\\		
		1162715& 0.79560&0.28227&0.01350&6.625 & 1.1598&1.949&6198\\		
		1163579&  0.74303&0.27985&0.01518&5.429 &1.1766 &1.38&6033\\
		\hline
	\end{tabular}
	\label{tab:main_res}
\end{table*}

\begin{figure}
 	\centering
 	\includegraphics[width=0.48\textwidth]{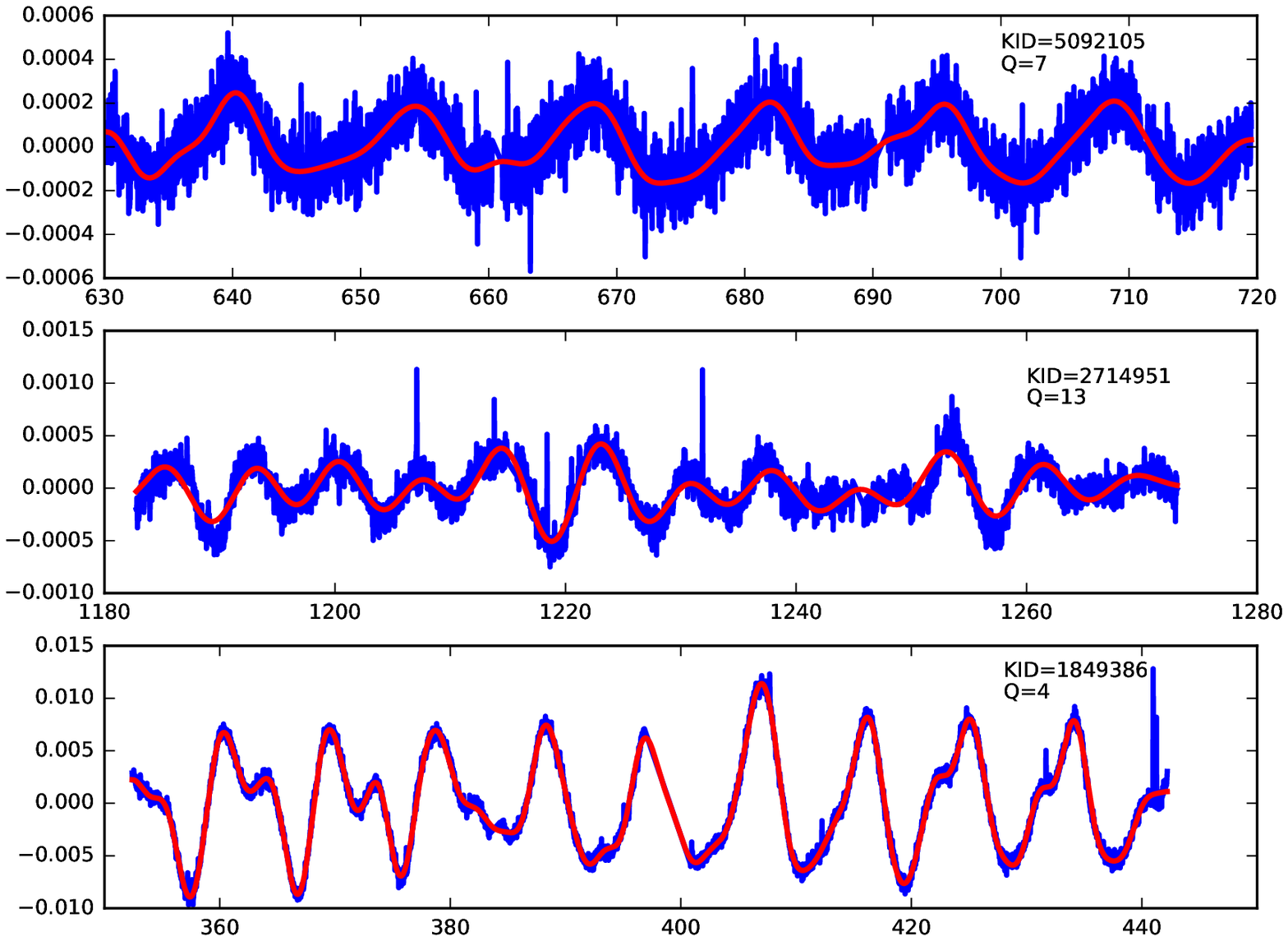}
 	\caption{An example of the \emph{Kepler} light curves with negative, zero, and positive correlations of $i_{\rm AC}$ and $R_{\rm eff}$ (from top to bottom). The original light curves are plotted in blue and the filtered light curves are in red. The associated \emph{Kepler} ID and quarter number are given in each panel.}
 	\label{fig:light}
 \end{figure}

According to Fig.(\ref{fig:cor_his}) the negative correlation stars are rare in our sample and correlation has a peak around $\sim 0.5$, so for most stars the two proxies are positively correlated. In Fig.(\ref{fig:light}), an example of the negative, zero, and positive correlation light curves is shown. The \emph{Kepler} ID and relevant quarter number are given in each panel. While these three \emph{Kepler} objects are selected randomly in our sample, there is an interesting feature in these light curves: the positive correlation star (bottom panel) fluctuates with relatively larger magnitude than the zero correlation star (middle panel), and also the zero correlation star's fluctuation magnitude is larger than the negative correlation star (upper panel). In the following, this property will be investigated statistically for all stars in our sample.

\begin{figure}
	\centering
	\includegraphics[width=0.48\textwidth]{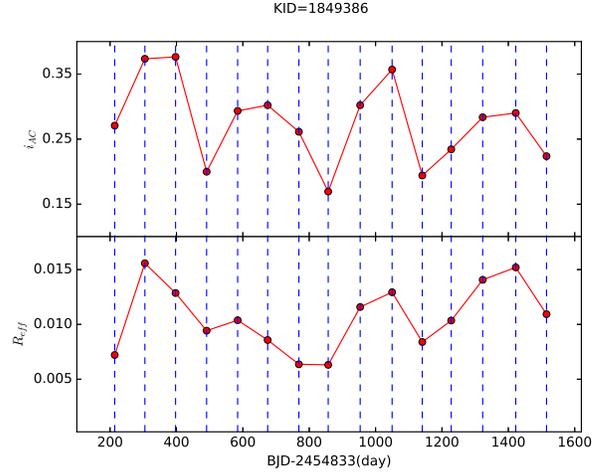}
	\caption{Variations of $i_{\rm AC}$ and $R_{\rm eff}$ for a positive correlation star. The \emph{Kepler} ID is shown at the top.}
	\label{fig:pos_cor}
\end{figure}

\begin{figure}
	\centering
	\includegraphics[width=0.48\textwidth]{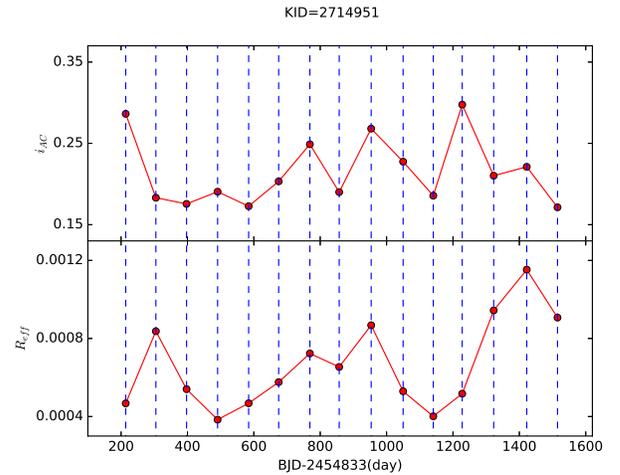}
	\caption{Variations of $i_{\rm AC}$ and $R_{\rm eff}$ for a zero correlation star. The \emph{Kepler} ID is shown at the top.}
	\label{fig:zer_cor}
\end{figure}

\begin{figure}
	\centering
	\includegraphics[width=0.48\textwidth]{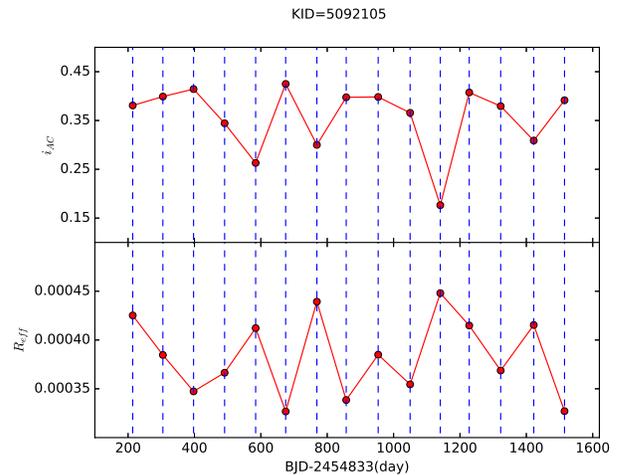}
	\caption{Variations of $i_{\rm AC}$ and $R_{\rm eff}$ for a negative correlation star. The \emph{Kepler} ID is shown at the top.}
	\label{fig:neg_cor}
\end{figure}

As shown in Figs.(\ref{fig:cor_his}) and (\ref{fig:light}), the correlation between the two magnetic proxies can be positive, zero or negative. Positive correlation means $i_{\rm AC}$ and $R_{\rm eff}$ evolve like each other: when one quantity increases, another also does, and vise versa. In other words, variations of $i_{\rm AC}$ and $R_{\rm eff}$ with time for the positive correlation stars are in the same phase. In Fig.(\ref{fig:pos_cor}), the time variations of $i_{\rm AC}$ and $R_{\rm eff}$ of the positive correlation star in Fig.(\ref{fig:light}) are shown. Clearly the variations of the two proxies are in the same phase. For the stars with zero correlation, the variations of $i_{\rm AC}$ and $R_{\rm eff}$ do not show any consistent behaviors. In Fig.(\ref{fig:zer_cor}), we show a typical star with zero correlation between the two quantities (the same zero correlation star in Fig.(\ref{fig:light})). For the stars with negative correlation, the two proxies vary almost inversely. That is, when $R_{\rm eff}$ increases $i_{\rm AC}$ decreases and vise versa. Our sample contains a small fraction of this kind of stars and a typical star with negative correlation is shown in Fig.(\ref{fig:neg_cor})(the same negative correlation star in Fig.(\ref{fig:light})). An illustrative example for the negative correlation star is the Sun. \cite{2015ApJS..221...18H} used the solar irradiance data observed by the SOHO spacecraft \citep{1995SoPh..162....1D} to obtain the correlation of the two proxies for the Sun. The analysis of solar irradiance between 1996 to 2009 yields Cor=-0.29 which indicates that at solar minimum, when $R_{\rm eff}$ is small, $i_{\rm AC}$ is relatively large. From a careful analysis of the solar light-curve at minimum \citep{2015ApJS..221...18H} showed that  the magnetic feature that causes the steady periodic light-curve fluctuation (corresponding to relatively larger $i_{\rm AC}$) at solar minimum is faculae (associated with enhanced network magnetic field) rather than sunspots (associated with intense and concentrated magnetic field).

\begin{figure}
	\centering
	\includegraphics[width=0.5\textwidth]{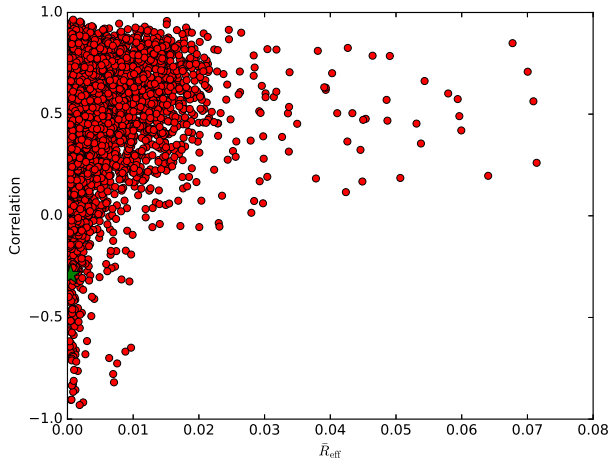}
	\caption{ Correlation versus $\bar{R}_{\rm eff}$ for all the stars in our sample. The green star symbol shows the position of the Sun for reference.}
	\label{fig:m_r-cor}
\end{figure}

As we mentioned above, one interesting fact which can be seen from Figs.(\ref{fig:pos_cor}), (\ref{fig:zer_cor}), and (\ref{fig:neg_cor}) is that the mean value of $R_{\rm eff}$ (denoted by $\bar{R}_{\rm eff}$) for a negative correlation star is smaller than that of the other two cases. This can happen by chance or be an essential physical property. To investigate this issue, in Fig.(\ref{fig:m_r-cor}) the scatter plot of correlation versus $\bar{R}_{\rm eff}$ for all the stars in our sample is depicted. The result shows that small values of $\bar{R}_{\rm eff}$ could happen on the negative, zero, and positive correlation stars, but negative correlation stars can't have large $\bar{R}_{\rm eff}$ value. Note that the Sun is a negative correlation star with $\bar{R}_{\rm eff}=0.00047$ \citep{2015ApJS..221...18H} and is indicated by a green star symbol in Fig.(\ref{fig:m_r-cor}).

\begin{figure}
	\centering
	\includegraphics[width=0.48\textwidth]{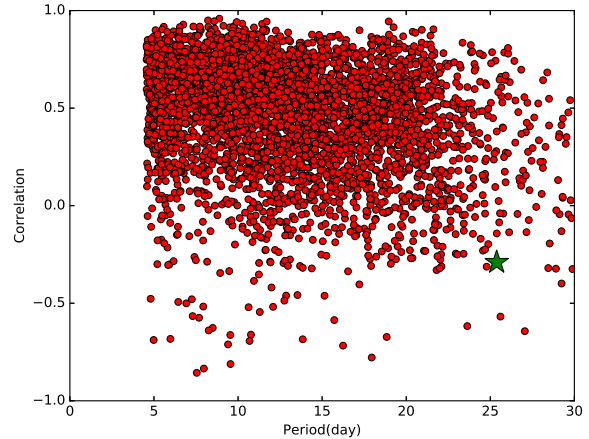}
	\caption{Correlation versus rotation period for all the stars in our sample. The green star symbol shows the position of the Sun for reference.}
	\label{fig:per}
\end{figure}

\begin{figure}
	\centering
	\includegraphics[width=0.48\textwidth]{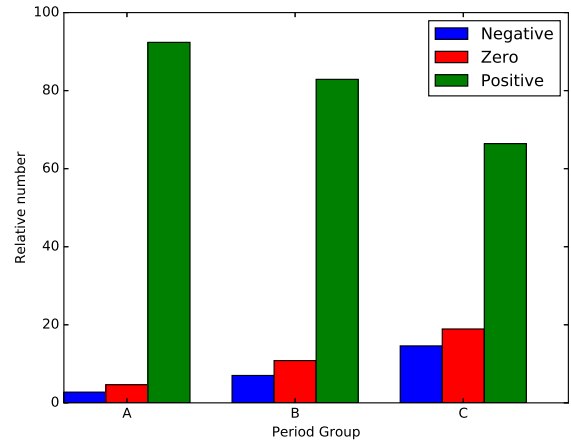}
	\caption{Relative numbers of positive, zero, and negative correlation stars in three period parts. A, B, and C indicate stars with period less than 10, between 10 and 20, and larger than 20 days, respectively.}
	\label{fig:per-num}
\end{figure}

Finally, in Fig.(\ref{fig:per}), the diagram of correlation versus rotation period for all the stars in our sample is shown. Same as Fig.(\ref{fig:m_r-cor}) the green star symbol shows the position of the Sun. Fig.(\ref{fig:per}) demonstrates a weak trend between the rotation period and the correlation of $i_{\rm AC}$ and $R_{\rm eff}$. That is, the correlation tends to be positive for stars with shorter period, and be zero and negative for stars with longer period. This relation has been proposed in the paper by \citep{2015ApJS..221...18H} and is confirmed by the statistical result in this work based on a larger sample of G-type stars. In support of this idea, we divide the stars in our sample into three parts, part A with period less than 10 days, part B with period between 10--20 days, and part C with period larger than 20 days. In each part we count the number of stars with positive, zero, and negative correlation and divide them by the total number of stars in each part. These relative numbers of stars are shown in Fig.(\ref{fig:per-num}). Clearly by increasing period, while number of stars with positive correlation is decreasing, number of negative correlation stars is increasing. In addition, this plot confirms that in each part, the two magnetic proxies are positively correlated for most of the stars.

\section{conclusion}\label{sec_concl}
By analyzing the correlation between the two light-curve-based magnetic proxies $i_{\rm AC}$ and $R_{\rm eff}$ for a large sample of G-type \emph{Kepler} targets, it is found that: (1) The two magnetic proxies are positively correlated for most of the stars in our sample, and the percentages of the negative, zero, and positive correlation stars are $4.27\%$, $6.81\%$, and $88.91\%$, respectively; (2) Negative correlation stars cannot have large magnitude of light-curve variation; (3) With the increase of rotation period, the relative number of positive correlation stars decreases and the negative correlation one increases.

 Based on above analysis results, it can be concluded that the stars with shorter rotation period tend to have positive correlation between the two proxies (see Figs.(\ref{fig:per}) and (\ref{fig:per-num})). From Fig.(\ref{fig:m_r-cor}) it can be further concluded that a good portion of the positive correlation stars have larger magnitude of light-curve variation (and hence more intense magnetic activities) than the negative correlation stars.

 The study in this work demonstrates that categorizing stars by correlation of these two magnetic proxies is potentially very useful. To have a more in-depth understanding of the physical meaning of the correlation, more effort is needed to reveal extra information about magnetic activity of stars in our sample. The morphology of the light-curve fluctuation and the superflare signal in the light curves can be two possible approaches (see Fig.(\ref{fig:light}) for an example), which we postpone to future works.

To end, we should note that performing our analysis for the smaller, lower surface temperature, M-type stars in \emph{Kepler} targets, is another possibility. A comparison of the results between these two categories might reveal extra information about magnetic activities of the stars. Investigation of magnetic activity of M-type stars, through two proxies used in this paper, is also in progress and will be published soon.

\acknowledgments
This paper includes data collected by the \emph{Kepler} mission which are downloaded from the Mikulski Archive for Space
Telescopes (MAST). Funding for the Kepler Mission is provided by
NASA Science Mission Directorate. The authors of this paper
gratefully acknowledge the entire Kepler team and all people who
contribute to the Kepler mission which provides us the opportunity to conduct this study.

\bibliographystyle{aasjournal}
\bibliography{ref}

\begin{thebibliography}{}
\expandafter\ifx\csname natexlab\endcsname\relax\def\natexlab#1{#1}\fi

\bibitem[{{Balona}(2015)}]{2015MNRAS.447.2714B}
{Balona}, L.~A. 2015, \mnras, 447, 2714

\bibitem[{{Basri} {et~al.}(2013){Basri}, {Walkowicz}, \&
  {Reiners}}]{2013ApJ...769...37B}
{Basri}, G., {Walkowicz}, L.~M., \& {Reiners}, A. 2013, \apj, 769, 37

\bibitem[{{Basri} {et~al.}(2011){Basri}, {Walkowicz}, {Batalha}, {Gilliland},
  {Jenkins}, {Borucki}, {Koch}, {Caldwell}, {Dupree}, {Latham}, {Marcy},
  {Meibom}, \& {Brown}}]{2011AJ....141...20B}
{Basri}, G., {Walkowicz}, L.~M., {Batalha}, N., {et~al.} 2011, \aj, 141, 20

\bibitem[{{Berdyugina}(2005)}]{2005LRSP....2....8B}
{Berdyugina}, S.~V. 2005, Living Reviews in Solar Physics, 2, 8

\bibitem[{{Borucki} {et~al.}(2010){Borucki}, {Koch}, {Basri}, {Batalha},
  {Brown}, {Caldwell}, {Caldwell}, {Christensen-Dalsgaard}, {Cochran},
  {DeVore}, {Dunham}, {Dupree}, {Gautier}, {Geary}, {Gilliland}, {Gould},
  {Howell}, {Jenkins}, {Kondo}, {Latham}, {Marcy}, {Meibom}, {Kjeldsen},
  {Lissauer}, {Monet}, {Morrison}, {Sasselov}, {Tarter}, {Boss}, {Brownlee},
  {Owen}, {Buzasi}, {Charbonneau}, {Doyle}, {Fortney}, {Ford}, {Holman},
  {Seager}, {Steffen}, {Welsh}, {Rowe}, {Anderson}, {Buchhave}, {Ciardi},
  {Walkowicz}, {Sherry}, {Horch}, {Isaacson}, {Everett}, {Fischer}, {Torres},
  {Johnson}, {Endl}, {MacQueen}, {Bryson}, {Dotson}, {Haas}, {Kolodziejczak},
  {Van Cleve}, {Chandrasekaran}, {Twicken}, {Quintana}, {Clarke}, {Allen},
  {Li}, {Wu}, {Tenenbaum}, {Verner}, {Bruhweiler}, {Barnes}, \&
  {Prsa}}]{2010Sci...327..977B}
{Borucki}, W.~J., {Koch}, D., {Basri}, G., {et~al.} 2010, Science, 327, 977

\bibitem[{{Brown} {et~al.}(2011){Brown}, {Latham}, {Everett}, \&
  {Esquerdo}}]{2011AJ....142..112B}
{Brown}, T.~M., {Latham}, D.~W., {Everett}, M.~E., \& {Esquerdo}, G.~A. 2011,
  \aj, 142, 112

\bibitem[{{Chaplin} {et~al.}(2011){Chaplin}, {Bedding}, {Bonanno}, {Broomhall},
  {Garc{\'{\i}}a}, {Hekker}, {Huber}, {Verner}, {Basu}, {Elsworth}, {Houdek},
  {Mathur}, {Mosser}, {New}, {Stevens}, {Appourchaux}, {Karoff}, {Metcalfe},
  {Molenda-{\.Z}akowicz}, {Monteiro}, {Thompson}, {Christensen-Dalsgaard},
  {Gilliland}, {Kawaler}, {Kjeldsen}, {Ballot}, {Benomar}, {Corsaro},
  {Campante}, {Gaulme}, {Hale}, {Handberg}, {Jarvis}, {R{\'e}gulo}, {Roxburgh},
  {Salabert}, {Stello}, {Mullally}, {Li}, \& {Wohler}}]{2011ApJ...732L...5C}
{Chaplin}, W.~J., {Bedding}, T.~R., {Bonanno}, A., {et~al.} 2011, \apjl, 732,
  L5

\bibitem[{{Cranmer} {et~al.}(2014){Cranmer}, {Bastien}, {Stassun}, \&
  {Saar}}]{2014ApJ...781..124C}
{Cranmer}, S.~R., {Bastien}, F.~A., {Stassun}, K.~G., \& {Saar}, S.~H. 2014,
  \apj, 781, 124

\bibitem[{{Davenport}(2016)}]{2016ApJ...829...23D}
{Davenport}, J.~R.~A. 2016, \apj, 829, 23

\bibitem[{{Debosscher} {et~al.}(2011){Debosscher}, {Blomme}, {Aerts}, \& {De
  Ridder}}]{2011A&A...529A..89D}
{Debosscher}, J., {Blomme}, J., {Aerts}, C., \& {De Ridder}, J. 2011, \aap,
  529, A89

\bibitem[{{Domingo} {et~al.}(1995){Domingo}, {Fleck}, \&
  {Poland}}]{1995SoPh..162....1D}
{Domingo}, V., {Fleck}, B., \& {Poland}, A.~I. 1995, \solphys, 162, 1

\bibitem[{{Fligge} {et~al.}(2000){Fligge}, {Solanki}, \&
  {Unruh}}]{2000A&A...353..380F}
{Fligge}, M., {Solanki}, S.~K., \& {Unruh}, Y.~C. 2000, \aap, 353, 380

\bibitem[{{Fr{\"o}hlich} \& {Lean}(2004)}]{2004A&ARv..12..273F}
{Fr{\"o}hlich}, C., \& {Lean}, J. 2004, \aapr, 12, 273

\bibitem[{{Garc{\'{\i}}a} {et~al.}(2010){Garc{\'{\i}}a}, {Mathur}, {Salabert},
  {Ballot}, {R{\'e}gulo}, {Metcalfe}, \& {Baglin}}]{2010Sci...329.1032G}
{Garc{\'{\i}}a}, R.~A., {Mathur}, S., {Salabert}, D., {et~al.} 2010, Science,
  329, 1032

\bibitem[{Gray(2005)}]{Gray-2005}
Gray, D. 2005, The Observation and Analysis of Stellar Photospheres (Cambridge
  University Press)

\bibitem[{{Haas} {et~al.}(2010){Haas}, {Batalha}, {Bryson}, {Caldwell},
  {Dotson}, {Hall}, {Jenkins}, {Klaus}, {Koch}, {Kolodziejczak}, {Middour},
  {Smith}, {Sobeck}, {Stober}, {Thompson}, \& {Van
  Cleve}}]{2010ApJ...713L.115H}
{Haas}, M.~R., {Batalha}, N.~M., {Bryson}, S.~T., {et~al.} 2010, \apjl, 713,
  L115

\bibitem[{{He} {et~al.}(2015){He}, {Wang}, \& {Yun}}]{2015ApJS..221...18H}
{He}, H., {Wang}, H., \& {Yun}, D. 2015, \apjs, 221, 18

\bibitem[{{Jenkins} {et~al.}(2010){Jenkins}, {Caldwell}, {Chandrasekaran},
  {Twicken}, {Bryson}, {Quintana}, {Clarke}, {Li}, {Allen}, {Tenenbaum}, {Wu},
  {Klaus}, {Van Cleve}, {Dotson}, {Haas}, {Gilliland}, {Koch}, \&
  {Borucki}}]{2010ApJ...713L.120J}
{Jenkins}, J.~M., {Caldwell}, D.~A., {Chandrasekaran}, H., {et~al.} 2010,
  \apjl, 713, L120

\bibitem[{{Jester} {et~al.}(2005){Jester}, {Schneider}, {Richards}, {Green},
  {Schmidt}, {Hall}, {Strauss}, {Vanden Berk}, {Stoughton}, {Gunn},
  {Brinkmann}, {Kent}, {Smith}, {Tucker}, \& {Yanny}}]{2005AJ....130..873J}
{Jester}, S., {Schneider}, D.~P., {Richards}, G.~T., {et~al.} 2005, \aj, 130,
  873

\bibitem[{{Kallinger} {et~al.}(2014){Kallinger}, {De Ridder}, {Hekker},
  {Mathur}, {Mosser}, {Gruberbauer}, {Garc{\'{\i}}a}, {Karoff}, \&
  {Ballot}}]{2014AA...570A..41K}
{Kallinger}, T., {De Ridder}, J., {Hekker}, S., {et~al.} 2014, \aap, 570, A41

\bibitem[{{Koch} {et~al.}(2010){Koch}, {Borucki}, {Basri}, {Batalha}, {Brown},
  {Caldwell}, {Christensen-Dalsgaard}, {Cochran}, {DeVore}, {Dunham},
  {Gautier}, {Geary}, {Gilliland}, {Gould}, {Jenkins}, {Kondo}, {Latham},
  {Lissauer}, {Marcy}, {Monet}, {Sasselov}, {Boss}, {Brownlee}, {Caldwell},
  {Dupree}, {Howell}, {Kjeldsen}, {Meibom}, {Morrison}, {Owen}, {Reitsema},
  {Tarter}, {Bryson}, {Dotson}, {Gazis}, {Haas}, {Kolodziejczak}, {Rowe}, {Van
  Cleve}, {Allen}, {Chandrasekaran}, {Clarke}, {Li}, {Quintana}, {Tenenbaum},
  {Twicken}, \& {Wu}}]{2010ApJ...713L..79K}
{Koch}, D.~G., {Borucki}, W.~J., {Basri}, G., {et~al.} 2010, \apjl, 713, L79

\bibitem[{{Lanza} {et~al.}(2003){Lanza}, {Rodon{\`o}}, {Pagano}, {Barge}, \&
  {Llebaria}}]{2003A&A...403.1135L}
{Lanza}, A.~F., {Rodon{\`o}}, M., {Pagano}, I., {Barge}, P., \& {Llebaria}, A.
  2003, \aap, 403, 1135

\bibitem[{{Lean} {et~al.}(1998){Lean}, {Cook}, {Marquette}, \&
  {Johannesson}}]{1998ApJ...492..390L}
{Lean}, J.~L., {Cook}, J., {Marquette}, W., \& {Johannesson}, A. 1998, \apj,
  492, 390

\bibitem[{{Maehara} {et~al.}(2012){Maehara}, {Shibayama}, {Notsu}, {Notsu},
  {Nagao}, {Kusaba}, {Honda}, {Nogami}, \& {Shibata}}]{2012Natur.485..478M}
{Maehara}, H., {Shibayama}, T., {Notsu}, S., {et~al.} 2012, \nat, 485, 478

\bibitem[{{McQuillan} {et~al.}(2013){McQuillan}, {Aigrain}, \&
  {Mazeh}}]{2013MNRAS.432.1203M}
{McQuillan}, A., {Aigrain}, S., \& {Mazeh}, T. 2013, \mnras, 432, 1203

\bibitem[{{McQuillan} {et~al.}(2014){McQuillan}, {Mazeh}, \&
  {Aigrain}}]{2014ApJS..211...24M}
{McQuillan}, A., {Mazeh}, T., \& {Aigrain}, S. 2014, \apjs, 211, 24

\bibitem[{{Nielsen} {et~al.}(2013){Nielsen}, {Gizon}, {Schunker}, \&
  {Karoff}}]{2013A&A...557L..10N}
{Nielsen}, M.~B., {Gizon}, L., {Schunker}, H., \& {Karoff}, C. 2013, \aap, 557,
  L10

\bibitem[{{Notsu} {et~al.}(2013){Notsu}, {Shibayama}, {Maehara}, {Notsu},
  {Nagao}, {Honda}, {Ishii}, {Nogami}, \& {Shibata}}]{2013ApJ...771..127N}
{Notsu}, Y., {Shibayama}, T., {Maehara}, H., {et~al.} 2013, \apj, 771, 127

\bibitem[{{Reiners}(2012)}]{2012LRSP....9....1R}
{Reiners}, A. 2012, Living Reviews in Solar Physics, 9, 1

\bibitem[{{Reinhold} {et~al.}(2013){Reinhold}, {Reiners}, \&
  {Basri}}]{2013A&A...560A...4R}
{Reinhold}, T., {Reiners}, A., \& {Basri}, G. 2013, \aap, 560, A4

\bibitem[{{Shibayama} {et~al.}(2013){Shibayama}, {Maehara}, {Notsu}, {Notsu},
  {Nagao}, {Honda}, {Ishii}, {Nogami}, \& {Shibata}}]{2013ApJS..209....5S}
{Shibayama}, T., {Maehara}, H., {Notsu}, S., {et~al.} 2013, \apjs, 209, 5

\bibitem[{{Smith} {et~al.}(2012){Smith}, {Stumpe}, {Van Cleve}, {Jenkins},
  {Barclay}, {Fanelli}, {Girouard}, {Kolodziejczak}, {McCauliff}, {Morris}, \&
  {Twicken}}]{2012PASP..124.1000S}
{Smith}, J.~C., {Stumpe}, M.~C., {Van Cleve}, J.~E., {et~al.} 2012, \pasp, 124,
  1000

\bibitem[{{Solanki} {et~al.}(2006){Solanki}, {Inhester}, \&
  {Sch{\"u}ssler}}]{2006RPPh...69..563S}
{Solanki}, S.~K., {Inhester}, B., \& {Sch{\"u}ssler}, M. 2006, Reports on
  Progress in Physics, 69, 563

\bibitem[{{Solanki} {et~al.}(2000){Solanki}, {Sch{\"u}ssler}, \&
  {Fligge}}]{2000Natur.408..445S}
{Solanki}, S.~K., {Sch{\"u}ssler}, M., \& {Fligge}, M. 2000, \nat, 408, 445

\bibitem[{{Stumpe} {et~al.}(2012){Stumpe}, {Smith}, {Van Cleve}, {Twicken},
  {Barclay}, {Fanelli}, {Girouard}, {Jenkins}, {Kolodziejczak}, {McCauliff}, \&
  {Morris}}]{2012PASP..124..985S}
{Stumpe}, M.~C., {Smith}, J.~C., {Van Cleve}, J.~E., {et~al.} 2012, \pasp, 124,
  985

\bibitem[{{Thompson} {et~al.}(2015){Thompson}, {Jenkins}, {Caldwell},
  {Barclay}, {Barentsen}, {Bryson}, {Burke}, {Campbell}, {Catanzarite},
  {Christiansen}, {Clarke}, {Colón}, {Cote}, {Coughlin}, {Girouard}, {Haas},
  {Ibrahim}, {Klaus}, {Li}, {McCauliff}, {Morris}, {Mullally}, {Rowe},
  {Sabale}, {Seader}, {Smith}, {Tenenbaum}, {Twicken}, {Uddin}, \&
  J.}]{Release25}
{Thompson}, S.~E., {Jenkins}, J.~M., {Caldwell}, D.~A., {et~al.} 2015, Kepler
  Data Release 25 Notes (KSCI-19065-001)

\bibitem[{{Walkowicz} {et~al.}(2011){Walkowicz}, {Basri}, {Batalha},
  {Gilliland}, {Jenkins}, {Borucki}, {Koch}, {Caldwell}, {Dupree}, {Latham},
  {Meibom}, {Howell}, {Brown}, \& {Bryson}}]{2011AJ....141...50W}
{Walkowicz}, L.~M., {Basri}, G., {Batalha}, N., {et~al.} 2011, \aj, 141, 50

\bibitem[{{Yun} {et~al.}(2016){Yun}, {Wang}, \& {He}}]{2016AcASn..57....9Y}
{Yun}, D., {Wang}, H.~N., \& {He}, H. 2016, Acta Astronomica Sinica, 57, 9

\bibitem[{{Zirin}(1985)}]{1985AuJPh..38..961Z}
{Zirin}, H. 1985, Australian Journal of Physics, 38, 961

\end{thebibliography}

\end{document}